# Penalized Split Criteria for Interpretable Trees

Alex Goldstein *
Department of Statistics; The Wharton School, University of Pennsylvania
and
Andreas Buja
Department of Statistics; The Wharton School, University of Pennsylvania

October 21, 2013

**Abstract**

This paper describes techniques for growing classification and regression trees designed to induce visually interpretable trees. This is achieved by penalizing splits that extend the subset of features used in a particular branch of the tree. After a brief motivation, we summarize existing methods and introduce new ones, providing illustrative examples throughout. Using a number of real classification and regression datasets, we find that these procedures can offer more interpretable fits than the CART methodology with very modest increases in out-of-sample loss.

## 1 Introduction

We assume familiarity with the techniques introduced in Breiman et al. (1984) for fitting binary trees to data. For brevity we refer to these techniques both collectively and individually by the acronym CART. Its authors state that CART is designed to "produce an accurate classifier or to uncover the predictive structure" of a problem. In comparison with the former task, the degree to which a model "uncovers structure" eludes quantification. We offer no help on this front, but adopt Breiman et al. (1984)'s preference for "simple characterizations of the conditions that

*The authors gratefully acknowledge support from the Simons Foundation Autism Research Initiative.

determine when an object is one class rather than another" as our guiding principle, which we call *interpretability*.

What is meant by a "simple characterization"? For classification trees, the question of whether we predict $y$ is one class or another is determined by the terminal node to which its associated $\boldsymbol{x}$ vector belongs. Hence the conditions leading to $y$'s predicted class are exactly the sequence of splitting rules that lead to its terminal node. As such, the tree that offers the simpler sequence of splits also offers the simpler explanation of $y$'s predicted class. In this sense, splitting procedures that encourage simple sequences of split rules can result in particularly interpretable trees. Such procedures are the focus of this paper.

In Section 2 we review the fundamentals of CART, paying special attention to gain and impurity, the critical functions for tree-growing. Further, we make the notion of "simple sequences of splits" more precise. In Section 3 we present novel tree growing techniques for the usual classification and regression settings where interpretability is desirable. Section 4 reviews the out-of-sample performance of these methods. The evidence suggests that in many cases the methods described in Section 3 yield interpretable trees with little sacrifice in generalization error. Section 5 concludes.

# 2 Fundamentals of Classification and Regression Trees

## 2.1 Splits and Splitting Criteria

Where possible we follow the terminology and notation of Breiman (1996b), as outlined below. Readers will recall that given a learning sample $\mathcal{L}$ of $N$ pairs $z_i = (y_i, \boldsymbol{x}_i)$ from an arbitrary distribution in which $\mathbb{E}(y|\boldsymbol{x}) = f(\boldsymbol{x})$, the algorithms described in Breiman et al. (1984) output a binary tree $\hat{f}(\boldsymbol{x})$ that aims to approximate $f$ or threshold $f(\boldsymbol{x})$ when $y$ is binary. Here $\hat{f}$ is called a classification or regression tree depending on whether $y$ is categorical or continuous, respectively.

For any $\boldsymbol{x}$, $\hat{f}(\boldsymbol{x})$ is given by the mean (for regression) or most common (in classification) $y_i$ value over all $i \in \mathcal{L}$ that are in the same terminal node as $\boldsymbol{x}$, denoted $\mathfrak{t}(\boldsymbol{x})$. In either case, all observations in a given node $\mathfrak{t}$ share the same fitted value, which we denote $j(\mathfrak{t})$ herein.

Each non-terminal node in the tree is defined by a *splitting rule* $s$. Each splitting rule comprises a pair $(x, t)$ consisting of a variable $x$ and a split location $t$. The rule $s = (x_1, 0)$, for instance, divides the $n_\mathfrak{t}$ observations in $\mathfrak{t}$ into two subsets, depending on whether each $\boldsymbol{x}$ vector has a positive first coordinate. In this example $s_x = x_1$ is termed the *split variable* and 0 the *split point*. The *growing* phase consists of selecting the best $s$ at $\mathfrak{t}$ and then sending $\mathfrak{t}$'s observations to the appropriate child nodes, where the recursion begins anew. Though the details of both growing and pruning certainly influence interpretability, our focus here is on tree-growing methods. Defining procedures for choosing splits that lead to interpretable trees is the subject of Section 3.

CART determines the "best $s$" by the *goodness of split criteria* or *gain* function $\theta(\mathfrak{t}, s)$ which quantifies the benefit of splitting node $\mathfrak{t}$ as per rule $s$. Each node splits at

$$s^\star = \underset{s \in \mathcal{S}}{\arg\max} \quad \theta(\mathfrak{t}, s), \tag{1}$$

meaning we choose the split that maximizes the split criterion, where $\mathcal{S}$ is the set of all possible splits including no split. For CART, $\theta$ is of the form

$$\theta(\mathfrak{t}, s) = \phi(\mathfrak{t}) - \left[\frac{n_{\mathfrak{t}_L}}{n_\mathfrak{t}}\phi(\mathfrak{t}_L) + \frac{n_{\mathfrak{t}_R}}{n_\mathfrak{t}}\phi(\mathfrak{t}_R)\right], \tag{2}$$

where $\mathfrak{t}_L$ and $\mathfrak{t}_R$ are the left and right child nodes defined by $s$, and $\phi$ is the loss or so-called *impurity* function. By multiplying $\phi_L$ and $\phi_R$ by the proportion of observations in the left and right child nodes, $\theta(\mathfrak{t}, s)$ measures the average improvement in impurity from splitting $\mathfrak{t}$ as per rule $s$. For convenience, Table 1 summarizes our notational conventions.

Table 1: Summary of notation

| Symbol | Definition |
|---|---|
| $\mathcal{L}$ | Training sample of $N$ $(y_i, \boldsymbol{x}_i)$ pairs |
| $\hat{f}$ | Recursive partitioning tree grown using the training sample |
| $\boldsymbol{x}$ | An arbitrary point in predictor space |
| $\hat{f}(\boldsymbol{x})$ | Tree $\hat{f}$'s fitted value at $\boldsymbol{x}$ |
| $\mathfrak{t}(\boldsymbol{x})$ | The terminal node to which $\boldsymbol{x}$ belongs |
| $j(\mathfrak{t})$ | The fitted value associated with node $\mathfrak{t}$ |
| $\mathfrak{t}_L, \mathfrak{t}_R$ | Node $\mathfrak{t}$'s left and right child nodes if $\mathfrak{t}$ is non-terminal |
| $n_\mathfrak{t}$ | Number of training observations in node $\mathfrak{t}$ |
| $s$ | Splitting rule consisting of a (split variable, split point) pair |
| $s_x$ | Split variable associated with splitting rule $s$ |
| $\theta$ | Goodness of split criteria / gain function |
| $\phi$ | Impurity function (see (1) above for the relation between $\theta$ and $\phi$) |
| $\hat{p}_{k,\mathfrak{t}}$ | Proportion of $y_i$'s in node $\mathfrak{t}$ that are of class $k$ (for categorical $y$) |
| $\Theta(\hat{f})$ | Loss function (MSE or misclassification rate), for use later |

## 2.2 CART Impurity Functions

In a regression setting we typically seek to minimize absolute or squared deviations between fitted and observed values. Though Breiman et al. (1984) presents regression trees based on both criteria, it is commonplace to use squared-error loss and so we set

$$\phi_R(\mathtt{t}) = \frac{1}{n_{\mathtt{t}}} \sum_{i \in \mathtt{t}} (y_i - j(\mathtt{t}))^2. \tag{3}$$

Recall that in regression we set $j(\mathtt{t})$ to the sample mean of the in-node $y$ values, and so readers will quickly identify (3) as $\mathtt{t}$'s (biased) sample variance, $\hat{\sigma}^2(\mathtt{t})$. Further, as the sample mean minimizes squared error loss, we see that $j(\mathtt{t})$ minimizes empirical within-node impurity.

Though intuitively appealing, when growing trees for classification we do *not* take $\phi$ to be the weighted average misclassification error (Breiman et al., 1984). The reason is that the misclassification rate is insensitive to certain distinctions in desirability of splits. As a heuristic example, consider the following proposed splits for classifying $y \in \{A, B\}$ in a 100 observation node with $n_A = 70$ and $n_B = 30$.

| Split | Left Node Distribution | Right Node Distribution |
|---|---|---|
| $s_1$ | $n_A = 45$, $n_B = 0$ | $n_A = 25$, $n_B = 30$ |
| $s_2$ | $n_A = 60$, $n_B = 15$ | $n_A = 10$, $n_B = 15$ |

Here $s_1$ and $s_2$ both have misclassification error of 0.25, even though $s_1$ yields a node without errors. Clearly $s_1$'s left node has zero impurity on $\mathcal{L}$ and requires no further splits, making $s_1$ preferable. The difficulty lies in the fact that the misclassification rate is piecewise linear in the sample proportion $p_A$, whereas the example illustrates that the impurity function should decrease more rapidly as $p_A \to 0$ or $p_A \to 1$. See Buja and Lee (2001) or Buja et al. (2005) for a more complete discussion of impurity functions for classification trees.

Instead, it is common to use either the *Gini* criterion or *Cross-entropy* criterion. For the $K$-class problem with $y \in \{1, 2, ...K\}$ the Gini criterion is written

$$\phi_G(\mathtt{t}) = \sum_{k \in K} \hat{p}_{k,\mathtt{t}}(1 - \hat{p}_{k,\mathtt{t}}) \tag{4}$$

where $\hat{p}_{k,\mathtt{t}}$ is the proportion of $y_i$'s in node $\mathtt{t}$'s that are of class $k$. Cross-entropy is defined

$$\phi_{CE}(\mathtt{t}) = \sum_{k \in K} \hat{p}_{k,\mathtt{t}} \log(\hat{p}_{k,\mathtt{t}}). \tag{5}$$

It is easy to verify that both functions satisfy the requirement above. Breiman (1996b) notes that empirically, Gini tends to yield splits resulting in purer nodes, especially when $K > 2$. In addition, if in-node sample proportions are interpreted as class probability estimates, Gini corresponds to squared-error loss (see Breiman et al. (1984) or Hastie et al. (2009)). In their informative description of the R package `rpart`, a popular implementation of CART, Therneau and Atkinson (1997) comment that from a practical perspective there is usually little difference between the methods, especially when $K = 2$. Like `rpart`, many software packages implement both criteria but default to Gini. For brevity we do likewise; when referring to the conventional method of growing classification trees we assume Gini impurity as defined in (4).

## 2.3 Interpretability of Trees

The interpretability of a particular tree is a function of its splitting rules. As an example, consider the regression tree in Figure 1. This tree, $\hat{f}$, is the result of applying the CART procedure to the Boston Housing data, where the goal is to fit median housing prices in census tracts using a variety of features about homes' average physical characteristics and locations. As our focus is on the growing phase rather than pruning, unless noted otherwise all trees herein cease splitting once the current node contains 5% of all observations.

Figure 1: CART fit to the Boston Housing Data. Terminal nodes are restricted to contain no fewer than 5% of all observations. In-sample $R^2 = 0.8$.

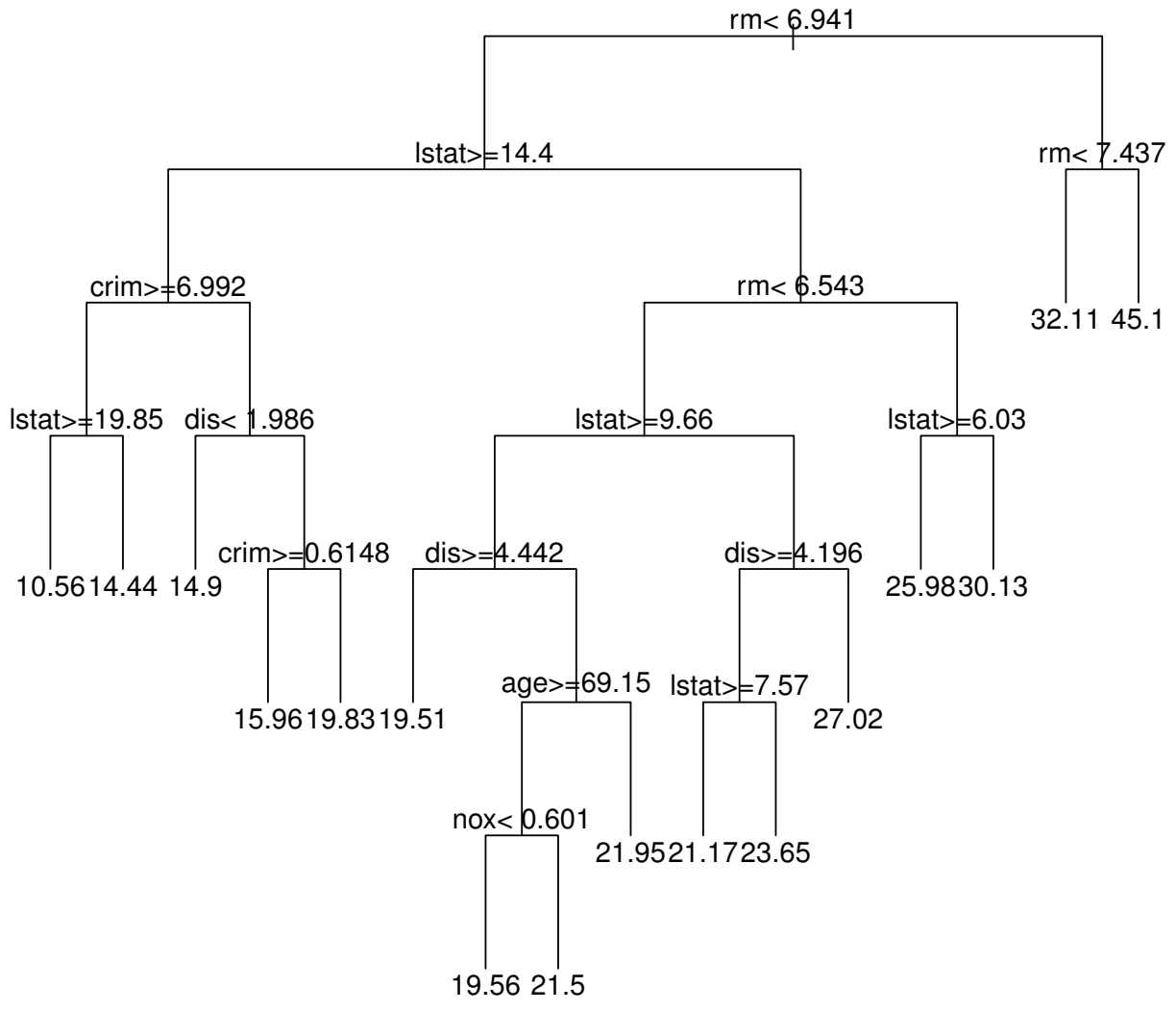

Now because $\hat{f}$ is a binary tree we can find $\hat{f}(\boldsymbol{x})$ simply by applying a series of rules. We write $\mathfrak{B}_\mathfrak{t}$ to denote the sequence of split variables leading to node $\mathfrak{t}$. Let $\mathfrak{t}$ be the left-most terminal node. Then the sequence of splits leading to $\mathfrak{t}$ is therefore $(\texttt{rm}, 6.941)$, $(\texttt{lstat}, 14.4)$, $(\texttt{crim}, 6.992)$, $(\texttt{lstat}, 19.85)$ corresponding to $\mathfrak{B}_\mathfrak{t} = \{\texttt{rm}, \texttt{lstat}, \texttt{crim}, \texttt{lstat}\}$. As noted previously, the fitted value of an observation for which $\boldsymbol{x} \in \mathfrak{t}$ is explained by simply enumerating this sequence of rules:

> "If `rm` is less than 6.94, `lstat` is greater than 14.4, `crim` is greater than 6.99, and `lstat` is greater than 19.85, then the fitted value is 10.56."

The node is at depth four and so the explanation is an intersection of four rules. Now clearly the more features used to reach a given terminal node `t`, the more difficult it is to summarize the partition of $\mathcal{X}$ `t` describes. Note, however, that in this case the explanation can be simplified by condensing the two statements about `lstat` into the single rule "`lstat` greater than 19.85." Similarly, the right-most terminal node can be described with the single rule "if `rm` exceeds 7.437, then $\hat{f}(\boldsymbol{x})$ is 45.1," despite the fact that it is at depth two.

More generally, because terminal nodes represent contiguous regions of $\mathcal{X}$, depth $d$ terminal nodes whose branches split on fewer than $d$ separate predictors can be interpreted as the intersection of fewer than $d$ rules. Put differently, sequential splits on the same variable are easily explained because they predict $y$ using a single dimension of $\mathcal{X}$. In the most extreme case, therefore, a node whose branch uses only a single variable corresponds to a contiguous region in $\mathcal{X}$ defined by a single dimension. This yields a single-rule explanation of the fitted value, regardless of the depth at which the node appears. Additionally, if these sequential splits uncover a monotonic relationship between the split points and fitted values, the explanation becomes easier still. In this sense, Breiman's concept of "simple characterizations" of $\mathcal{X}$ can be understood in part by the extent to which a tree's branches tend to reuse split variables.

# 3 Penalized Split Criteria for Interpretable Trees

## 3.1 Penalized Split Criteria

As we have seen, branches comprising small subsets of predictors are more interpretable than those containing new predictors at each split point. With this in mind, the criterion presented in this section encourages interpretable trees by penalizing splits that extend the set of features used in a given branch. Under this criterion the chosen split $s^\star$ is not necessarily the one that most reduces impurity, which obviously worsens the extent to which the tree fits the data. Nevertheless, it is encouraging that the presence of a single split which minimizes impurity does not imply the absence of other suitable split options, even if minimizing impurity is the sole objective. Readers familiar with the literature will recall that the chosen split $s^\star$ can be quite unstable, and that in reality many different splits may result in similar values of the gain criteria. In Breiman et al. (1984) the authors describe this phenomenon as follows.

> *At any given node, there may be a number of splits on different variables, all of which give almost the same decrease in impurity. Since the data are noisy, the choice between competing splits is almost random.*

As pointed out by many authors, the variability of CART splits is a drawback from a bias-variance perspective (see Breiman (1996a) and Hastie et al. (2009)). Here we focus on interpretability, and in the following sections we show how the presence of multiple splits with similar $\phi$ values can actually be advantageous for growing interpretable trees.

The central idea is that if choosing a particular split rule from a set of competing rules with similar $\phi$'s is "almost random" as Breiman et al. (1984) asserts, then selecting the most interpretable one from the set rather than that which strictly maximizes the gain function should yield a tree that both fits the data and is easy to explain. To that end, given a non-negative *penalty function* $\gamma$ for splitting $\mathfrak{t}$ as per rule $s$, we split according to

$$s^\star = \underset{s \in \mathcal{S}}{\arg\max} \quad \left[\theta(\mathfrak{t}, s) - \gamma_k(\mathfrak{t}, s, \mathfrak{B}_\mathfrak{t})\right], \tag{6}$$

where the $k$ refers to a penalization constant to be discussed shortly. As before, $\mathfrak{B}_\mathfrak{t}$ is the ordered list of split variables used in the branch of the tree leading to $\mathfrak{t}$. The algorithm is still recursive but is now path dependent. Particular definitions of $\gamma$ are the subject of Sections 3.2 and 3.3.

The constant $k$ will be a tuning parameter that controls the tradeoff between the gain function and the penalty: high $k$ values will correspond to a strong preference for interpretable splits, potentially at the gain function's expense. Naturally, choosing splits with less than the maximal gain can result in reduced fit in terms of MSE or the misclassification rate. Nevertheless, as we shall see in the subsequent sections, in many cases the reduction is not drastic and could well be worth the improvement in interpretability. Of course the nature of the tradeoff varies with the dataset, and so it is advisable to run the algorithm for a variety of $k$ values. If we do not wish to use the tree for out-of-sample prediction this could very well be the end of the story – we simply choose the tree that yields the best combination of fit and interpretability for the problem at hand.

If a more systematic approach is desired, a natural procedure is to select the highest $k$ that results in a global fit no worse than that of the unpenalized tree's by some predefined fraction. We define this formally as follows. Recalling that $\mathcal{L}$ denotes our learning sample of $N$ $(y_i, \boldsymbol{x}_i)$ pairs, we write $\Theta[\hat{f}, \mathcal{L}]$ to denote tree $\hat{f}$'s loss evaluated on $\mathcal{L}$. At this point we only consider in-sample metrics (Section 4 discusses penalization's out-of-sample performance), and so $\mathcal{L}$ serves as $\hat{f}$'s training data as well. In regression, for example, we take

$$\Theta[\hat{f}, \mathcal{L}] = \sum_{i \in \mathcal{L}} (y_i - \hat{f}(\boldsymbol{x}_i))^2, \tag{7}$$

the usual squared-error loss. For convenience, in plots and tables we re-express this quantity as $R^2$ in order to remove the scale of $y$. In classification we let $\Theta$ be the misclassification rate (MR). Writing $\hat{f}_k$ to indicate a tree grown with a particular tuning parameter, we choose the parameter

$k^\star$ as per

$$k^\star = \arg\max_k \ \left\{ k : \Theta[\hat{f}_k, \mathcal{L}] \ \le \ (1+c)\ \Theta[\hat{f}_0, \mathcal{L}] \right\}, \tag{8}$$

where $c > 0$. That is, we choose the largest $k$ that still results in a tree whose loss is no worse than that of the unpenalized tree's by $100c\%$. Unless noted otherwise, all $k$'s for the penalized trees displayed in Sections 3.2 and 3.3 are chosen according to this procedure with $c = 0.10$.

Note that in regression $\theta(\mathfrak{t}, s)$ is the decrease in mean squared error, which is dependent on the scale of the response variable. Penalizing MSE directly means the choice of $k$ in (6) is dependent on the level of $y$ in a given problem. To make $k$ values comparable across datasets, in the sections below we re-express $\theta$ to measure the *proportional* improvement in impurity gained by splitting $\mathfrak{t}$ as per rule $s$. The details of the scaling vary with the impurity function and are deferred to Appendix A.2, but in each case we ensure that $\theta \le 1$ for all $s \in \mathcal{S}$, we prefer splits with larger $\theta$, and we are indifferent between splitting and not splitting when $\theta = 0$. Herein we assume scaled gain functions, letting us restrict $k$ to the interval $[0, 1]$.

## 3.2 New Variable Penalty

The first of our new methods is targeted at limiting the number of predictors used to reach a tree's terminal nodes. As we have described, the more variables used to reach $\mathfrak{t}$ the more complex the explanation of $\mathfrak{t}$'s subset of $\mathcal{X}$, and so in cases where many splits offer nearly the same $\phi$ it may be preferable to choose a split on a variable already used in $\mathfrak{B}_\mathfrak{t}$.

Letting $s_x \in \{1, ..., p\}$ denote rule $s$'s split variable, the *new variable penalty* is written

$$\gamma_k(\mathfrak{t}, s, \mathfrak{B}_\mathfrak{t}) = k\mathbf{1}(s_x \notin \mathfrak{B}_\mathfrak{t}). \tag{9}$$

Hence if $s$ introduces a new variable into the branch the penalty is $k$. If $s$ uses a previously used variable, there is no penalty. Thus splits that introduce new variables must improve $\theta$ by at least $k$ in order to be selected, whereas splits on old variables can be selected so long as the improvement is greater than 0. Whatever the split criteria, amongst many splits with similar $\theta$'s, using (9) gives preference to splits that do not introduce new variables into the branch. This penalty (and more generally any penalized criteria written in the form of Equation 6) can be made compatible with any suitably scaled split criteria. In the following we demonstrate the performance of the penalty (9) on the previously used datasets for a selection of split criteria.

In Figure 2 we compare trees grown to the Boston Housing data using (3), the conventional CART regression criteria, with and without the new variable penalty. First we note that despite the penalization, the $R^2$ values are comparable. The trees are equivalent up to the third level of splits, where the conventionally grown tree (Figure 2a) introduces `crim` into the leftmost branch.

All told, the unpenalized tree uses as many as five variables in reaching a terminal node, whereas the penalized tree (Figure 2b) never uses more than three. This makes a considerable difference when one attempts to explain the fit at a particular node. For instance, the region described by the bottom-left node of the penalized tree (for which $j(\mathtt{t}) = 20.63$) might be described by saying "if `rm` is less than 5.85 and `lstat` is between 11.69 and 14.4, the fitted value is 20.63." Constructing an analogous description of the bottom-left node of the unpenalized tree is substantially more tedious.

Figure 2: CART applied the Boston Housing Data with and without the New Variable Penalty.

(a) Unpenalized
In-sample $R^2 = 0.8$

(b) New Variable Penalty ($k^\star = 0.27$)
In-sample $R^2 = 0.79$

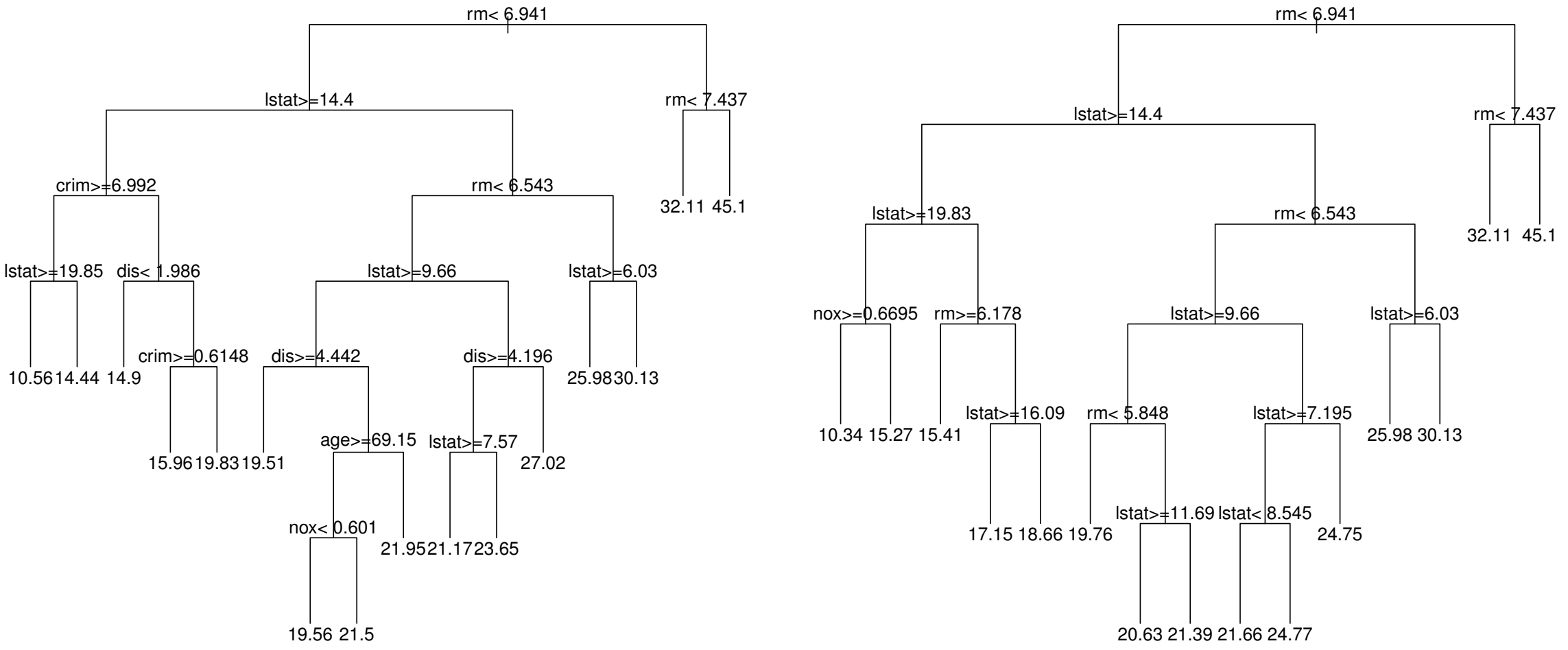

As per (8), 0.27 is the maximal value for $k$ that achieves a mean-squared error no more than 1.10 times that of the traditionally grown tree. Of course depending on how the analyst values fit versus interpretability, he can use higher values for $k$ resulting in even fewer variables used and a commensurate increase in in-sample MSE (decrease in $R^2$). For instance, Figure 3 uses $k = .4$, and largely describes the monotonic relationship between average home size and median prices.

Figure 3: CART fit to the Boston Housing Data with the New Variable Penalty ($k^{\star}$=0.4). In-sample $R^2$=0.67.

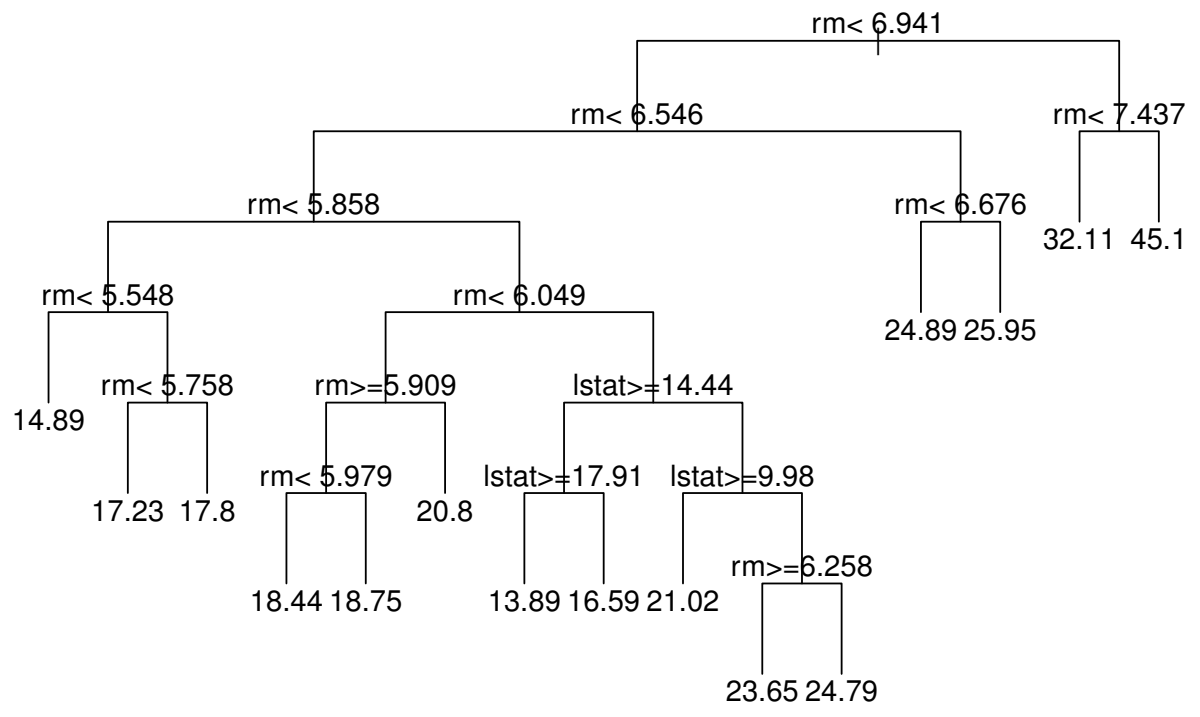

The penalization framework applies to split criteria besides the usual CART methodology. As an example, we consider the one-sided high means criteria described in Buja and Lee (2001). Unpenalized, this method chooses the split $s$ that isolates the single child node with the highest mean:

$$s^{\star}_{hm} = \arg\max_{s \in \mathcal{S}} \left\{ \max_{s} \left[ \bar{y}_{\mathrm{t}_L}, \bar{y}_{\mathrm{t}_R} \right] \right\}. \tag{10}$$

An overview of the one-sided procedures introduced in Buja and Lee (2001) is contained in Appendix A.1. Applying this procedure to the Boston Housing data yields the tree in Figure 4a, with the penalized version appearing in Figure 4b. The $R^2$ values are comparable, but the penalized tree is considerably simpler as it involves only three predictors instead of six. The trees are identical until the unpenalized tree splits on `dis`. Further down, the unpenalized tree splits on `nox` and `tax`, whereas the penalized tree uses only `crim` and `lstat`, leaving the monotonic relationships undisturbed.

Figure 4: One-sided High-Means fit to the Boston Housing Data.

(a) Unpenalized
In-sample $R^2 = 0.79$

(b) New Variable Penalty ($k^\star = 0.08$)
In-sample $R^2 = 0.78$

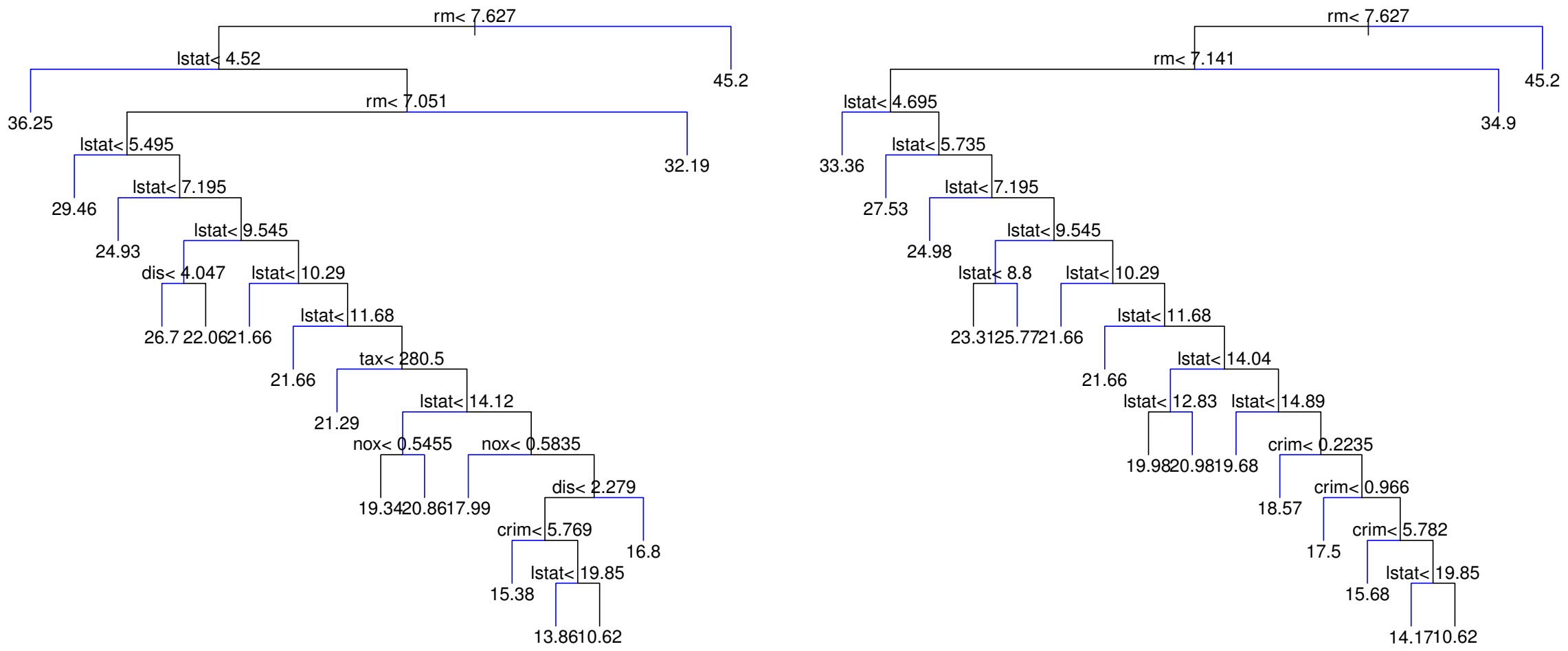

Turning to classification, Figure 5 combines the new variable penalty with Buja and Lee (2001)'s one-sided purity criteria, which splits so as to isolate the single child node with minimum Gini (minimum classification impurity). Comparing the penalized tree in Figure 5b with its unpenalized counterpart in Figure 5a, we see that we can achieve less than 10% increase in the in-sample misclassification rate while reducing the total number of predictors used from seven to two. Here applying the new variable penalty allows us to uncover high-purity regions of $\mathcal{X}$ that are also relatively simple to interpret.

## 3.3 EMA-Style Penalty

Let us consider more closely the four leaf nodes at depth 6 in the penalized tree in Figure 2b (the leftmost of these leaf nodes has $j(\mathtt{t}) = 20.63$). Using the new variable penalty allows us to see that the fits here depends on both `rm` and `lstat`. This represents an improvement over the corresponding branches in the unpenalized tree in Figure 2a that eventually split on `dis`, `age` and `nox`. Nevertheless, the fact that the predictors are interleaved makes constructing a more precise explanation difficult. Longer sequences of splits on the same variable would enable us to interpret the fits as monotonic relationships in `rm` and/or `lstat`, but here that is not possible. This should be no surprise – while (9) expresses our preference for using fewer variables, it is indifferent to the ordering of variables in a given branch.

Our second method targets both preferences. Here we penalize not only new variables, but also favor variables used recently in the branch. We achieve this by employing an *exponential moving*

Figure 5: One-Sided Purity fit to the Pima Indians Diabetes Data with and without the New Variable Penalty. (MR = Misclassification Rate)

(a) Unpenalized
In-sample MR = 0.21

(b) New Variable Penalty ($k^\star = 0.63$)
In-sample MR = 0.23

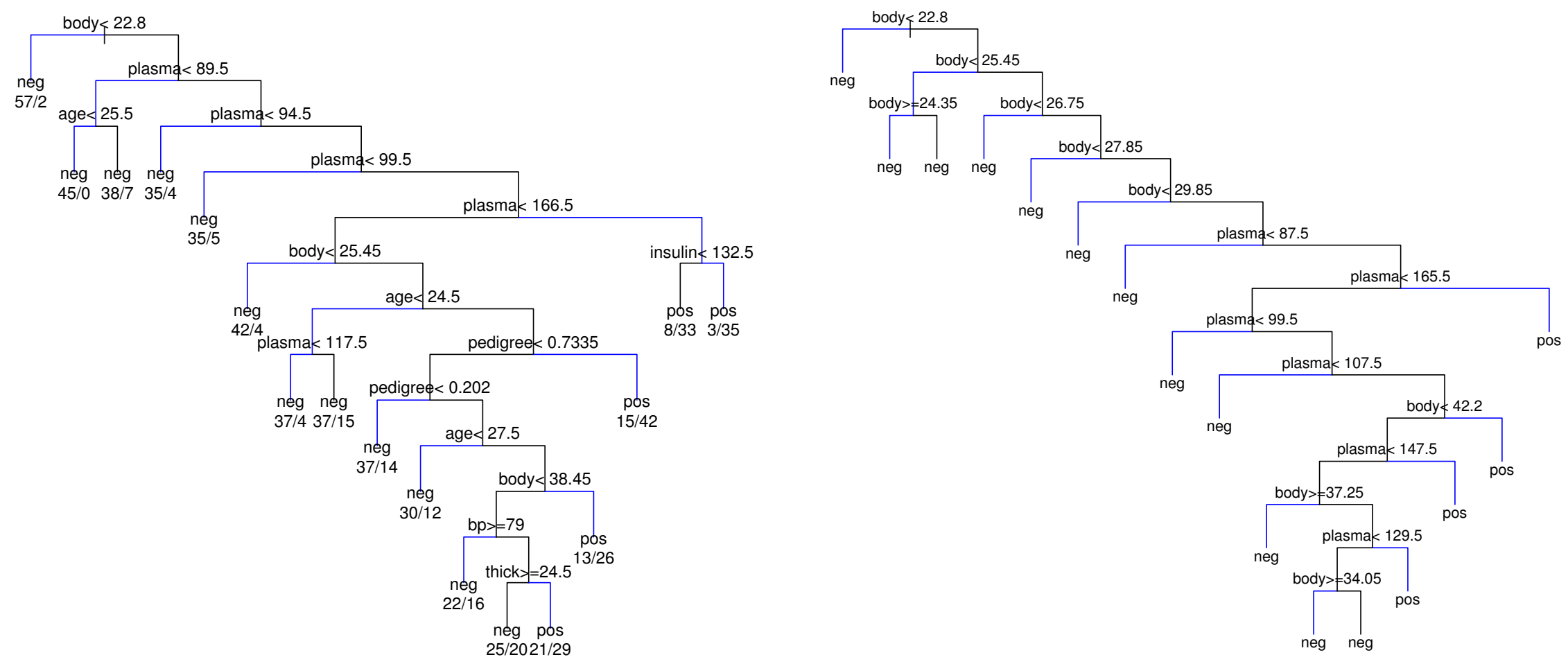

*average-style* penalty, defined as:

$$\gamma_k(\mathfrak{t}, s, \mathfrak{B}_\mathfrak{t}) = \sum_{j=0}^{d-1} \mathbf{1}(s_x \neq s_j) k(1-k)^{(d-1)-j} \quad \text{for } d > 0, \tag{11}$$

and otherwise 0. As before, $k \in [0, 1]$ is the user-specified penalty constant and $s_x$ is the variable corresponding to the proposed split $s$. We let $j \in \{0, 1, \ldots\}$ index the depth of $\mathfrak{B}_\mathfrak{t}$'s nodes, and so $s_j$ is the split variable in $\mathfrak{B}_\mathfrak{t}$ at depth $j$. The branches we last discussed from Figure 2 have $s_0$=`rm`, $s_1$=`lstat` and $s_2$=`rm`, for instance. Here $d$ is the depth of the branch not including the proposed split, or equivalently, the number of nodes in $\mathfrak{B}_\mathfrak{t}$. Hence when considering candidates for the second split in a branch we have $d = 1$. Obviously when considering the root split there should be no penalty (nor does (11) make sense), and so we set $\gamma = 0$.

Setting aside the notational details, we see that (11) is an exponential moving average of indicator functions. The $j$-th indicator is 1 if $s$'s split variable is different from the variable used at depth $j$. If $s$ splits on the same variable, as we prefer, the indicator is 0. Further, as $j \to 0$ we know $k(1-k)^{(d-1)-j}$ decreases, and so the weights attenuate as we move up $\mathfrak{B}_\mathfrak{t}$ towards the root. This conforms to our preferences: splitting a node on a different predictor from its parent is a graver offense than splitting on a different predictor from the root. Correspondingly, the former infraction contributes more to $\gamma$ than the latter. Lastly we note that setting $k = 0$ recovers the unpenalized version of the splitting criteria.

Figure 6 displays a regression tree grown using the CART procedure but with the EMA-style penalty. The unpenalized version of this tree appears in Figure 2a. Immediately we see that the new penalty eliminates the previously observed tendency for consecutive nodes to switch between splitting on `rm` and `lstat`. The benefit is that the fit is easily explained primarily in terms of two monotonic relationships: for areas with very large homes (`rm` $> 6.94$) prices are monotonically increasing in home size, and for the remaining areas prices are decreasing in `lstat`. A very similar story emerges from using the EMA penalty with the high-means criteria, as displayed in Figure 7. In fact, some of the nodes in these trees characterize the exact same partition of $\mathcal{X}$.

Figure 6: CART fit to the Boston Housing data with the EMA-Style penalty ($k^\star = .15$, In-sample $R^2 = 0.77$). In comparison with the unpenalized version in Figure 2, this tree uses only two predictors.

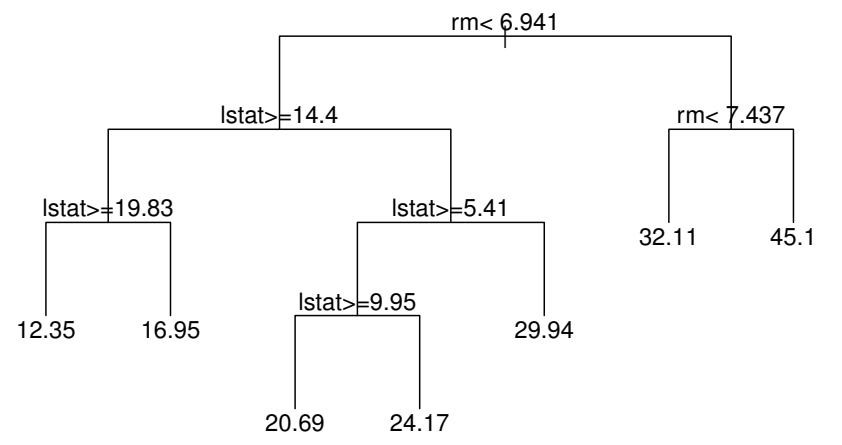

Figure 7: High-Means fit to the Boston Housing data with the EMA penalty ($k^\star$=.01). In-sample $R^2 = 0.78$.

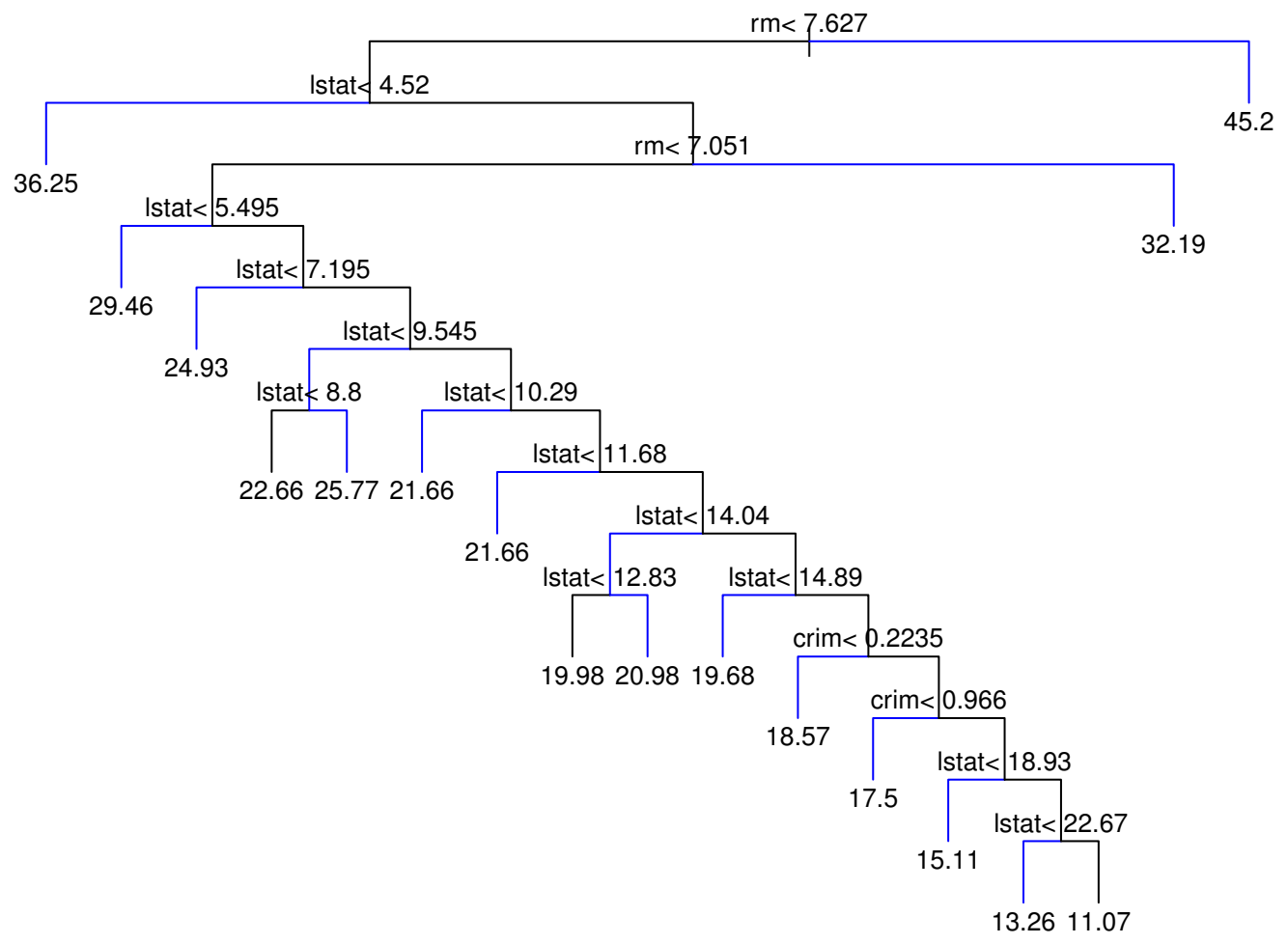

In Figure 8 we apply the the EMA penalty to the Pima Indians data using Buja and Lee (2001)'s one-sided extremes criterion. This procedure chooses the split that results in the single child node with the highest sample proportion of a specified class. Here we search for regions of $\mathcal{X}$ associated with high incidence of diabetes. From previous examples we know that this dataset can withstand very high penalties before the misclassification rate breaks down. Hence in this example we set $c$ to 0 and choose the penalization parameter whose associated tree's misclassification rate is *no higher* than the unpenalized version's. Notice that the trees have the same shape and misclassification rates, but the right tree uses only a single predictor. The unpenalized tree, in comparison, never uses the same variable more than twice consecutively and employs seven predictors in all. Figure 9 displays the one-sided purity tree with and without the EMA penalty when $c = 0.10$.

Figure 8: One-Sided Extremes fit to the Pima Indians data with and without the EMA penalty. Figure 8b uses the EMA penalty with the highest penalty parameter such that the penalized tree's misclassification is no higher than that of the unpenalized tree. Note the penalized tree uses only `plasma`, whereas the unpenalized tree uses 7 predictors.

(a) Unpenalized
In-sample MR = 0.25

(b) EMA Penalty ($k^\star = 0.70$)
In-sample MR = 0.25

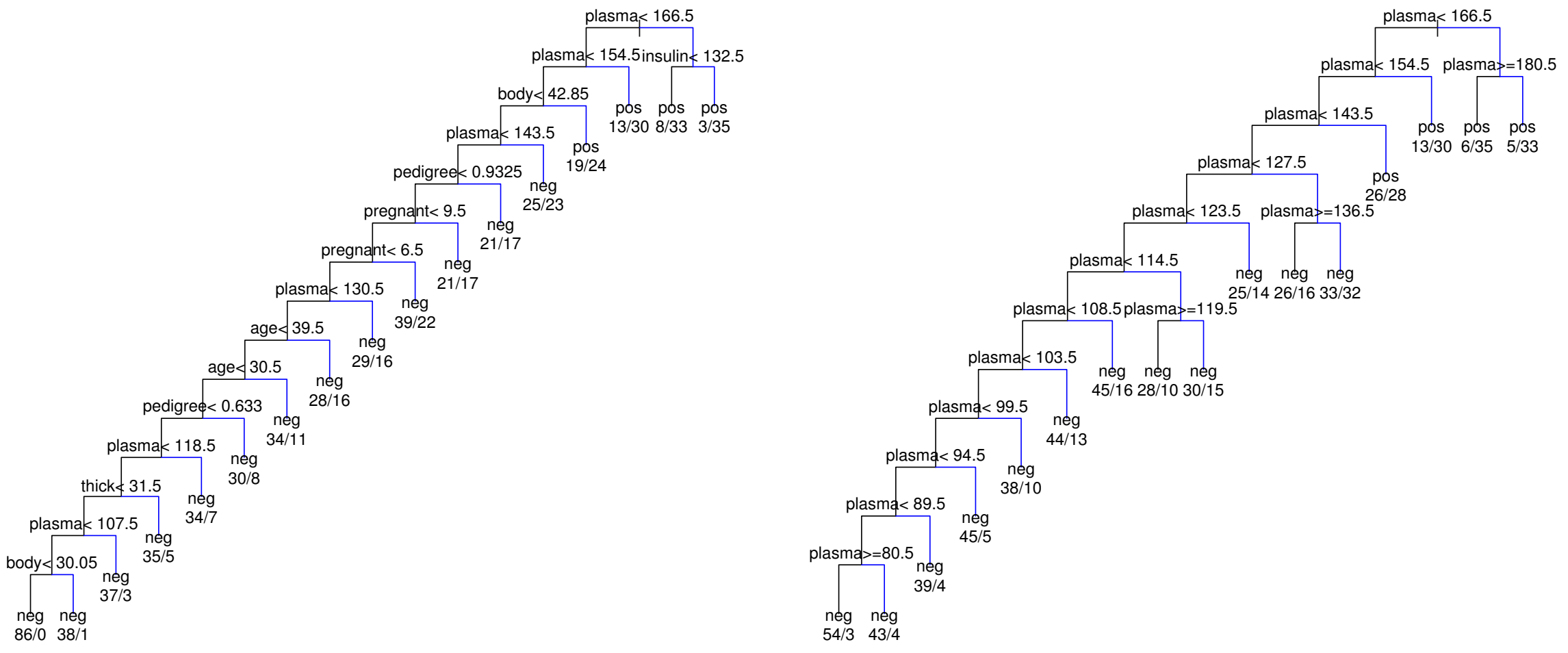

# 4 Out of Sample Performance

We have seen that one-sided split criteria and penalization often yield more interpretable trees than the traditional CART methodology with only modest sacrifices in in-sample loss, $\Theta$. Until now we have computed loss over our learning sample $\mathcal{L}$, but naturally it is important to understand how these techniques fare on new data, $z^{new} = (y^{new}, \boldsymbol{x}^{new})$, as well. To that end, in this section we study the impact of the various techniques for growing $\hat{f}$ on the risk, defined by

$$R = \int_{z^{new}} \int_{\mathcal{L}} \Theta[\hat{f}_{\mathcal{L}}, (y^{new}, \boldsymbol{x}^{new})] \, \mathrm{d}P(\mathcal{L}) \mathrm{d}P(z^{new}). \tag{12}$$

Figure 9: One-Sided Purity fit to the Pima Indians data with and without the EMA penalty.

(a) Unpenalized
In-sample MR = 0.21

(b) EMA Penalty ($k^{\star} = 0.03$)
In-sample MR = 0.23

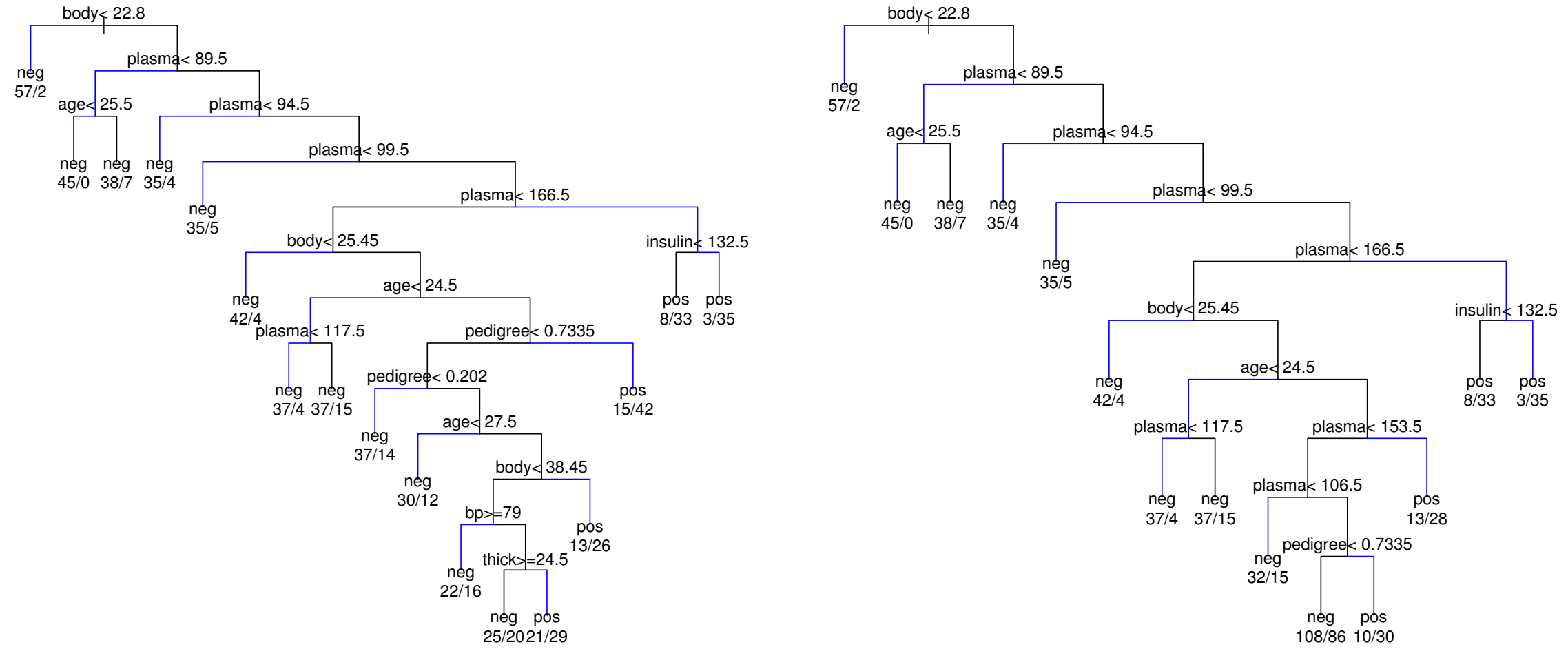

We write $\hat{f}_{\mathcal{L}}$ to emphasize that the fitted tree is a function of the training sample $\mathcal{L}$. In general our results suggest that applying an interpretability penalty to a given splitting criteria has very little impact on out-of-sample loss in comparison with the unpenalized criteria. This holds for both classification and regression problems over a variety of splitting methods. In the remainder of this section we discuss these results in greater detail.

As we have neither true distribution functions nor an elegant form for the fitting procedure $\mathcal{L} \rightarrow \hat{f}_{\mathcal{L}}$ at our disposal, we study (12) using the "out-of-bag" generalization error estimate discussed in Breiman (1997). For each dataset we take $B$ bootstrap samples $L_1, ..., L_B$ from $\mathcal{L}$. Let the bootstrap samples be indexed by $b \in \{1, \dots B\}$. Observations in $\mathcal{L}$ not in $L_b$ are set aside as holdout data, $H_b$. Using a tree fitting procedure $F$ we fit a tree to each sample. Then for each tree we evaluate loss $\Theta$ on its holdout data $H_b$, yielding an estimate of generalization error $\hat{\Theta}_b$. The procedure is given completely by Algorithm 1. We then approximate $R$ with the mean of the $\hat{\Theta}$ values:

$$R_{OOB} = \frac{1}{B} \sum_{b=1}^{B} \hat{\Theta}_b. \tag{13}$$

In Algorithm 1 $F$ represents the fitting procedure. Here we use both CART and the one-sided splitting criterion introduced in Buja and Lee (2001). The one-sided splitting criterion is written

$$\theta_{OS}(\mathfrak{t}, s) = \phi(\mathfrak{t}) - \min\left[\phi(\mathfrak{t}_L),\ \phi(\mathfrak{t}_R)\right], \tag{14}$$

**Input**: $\mathcal{L}$: learning sample of $N$ $(y_i, \boldsymbol{x}_i)$ pairs
$B$: number of bootstrap samples
$F$: tree-fitting procedure, including method for choosing $k^\star$
$\Theta[\hat{f}, (y, \boldsymbol{x})]$: function specifying loss from estimating $y$ with $\hat{f}(\boldsymbol{x})$

**Output**: $\hat{\Theta}$, a $B \times 1$ vector of the estimated risk for each bootstrap replicate

**Initialize**: $l \leftarrow \mathbf{0}_{B\times 1}$
$nl \leftarrow \mathbf{0}_{B\times 1}$
$\hat{\Theta} \leftarrow \mathbf{0}_{B\times 1}$

```
for b = 1 to B do
    # bootstrap sampling:
    L_b ← N    # sample of N observations with replacement from 𝓛;
    H_b ← 𝓛 \ L_b;

    # fit a tree to the bootstrap learning sample:
    T_b ← F(L_b);

    # evaluate the tree on holdout data:
    for i = 1 to N do
        if (y_i, x_i) ∈ H_b then
            λ ← Θ[T_b, (y_i, x_i)];
            l[b] ← l[b] + λ;
            nl[b] ← nl[b] + 1;
        end
    end
end

# normalize:
for i = b to B do
    Θ̂[b] ← l[b]/nl[b];
end

return Θ̂;
```

**Algorithm 1:** Procedure for generating out-of-bag error estimates.

with $s^\star \in \mathcal{S}$ still chosen by maximizing the gain function as in (1). In replacing (2)'s weighted sum over child nodes with minimization, (14) *favors* splits with low $\phi$ on the left at the expense of high $\phi$ on the right and vice versa, regardless of relative node size. Appendix A.1 describes how the high means and one-sided purity methodologies seen previously fit into this framework in addition to summarizing the remainder of the procedures described by Buja and Lee (2001).

Turning to the interpretability penalties, the reader will recall that we set the penalization constant $k$ using (8). Roughly speaking this procedure aims to return the most interpretable tree that still achieves a certain fraction of the unpenalized method's performance on the data at hand. When we write "penalization method" or "penalization procedure" we mean a particular penalization function coupled with our rule for choosing $k$. To study the penalties' out-of-sample performance, we compute the out-of-bag error estimate as before but apply (8) to each bootstrap learning sample. By this we mean that the $F$ from Algorithm 1's line $T_b \leftarrow F(L_b)$ includes the search over possible $k$ values. Hence $\hat{\Theta}$ remains a metric of out-of-sample performance.

Starting with Table 2 we display the estimated loss obtained from applying each splitting criteria and penalty method combination (including no penalty) to our datasets. We set $B = 100$, $\boldsymbol{k} = (0.01, 0.02, \ldots, 0.99)$, and use $c = .10$. For the penalized methods, the column entitled "Average $k^\star$" reports the mean $k$ value selected across the $B$ bootstrap samples. Low average $k^\star$ values suggest that on average, the splits chosen by the non-penalized methods have relatively few competitors in terms of reducing loss. The two wine datasets are examples of this – apparently in predicting wine quality, swapping the "best" split for a more interpretable one coincides with a substantial increase in MSE. In contrast, high $k^\star$, such as those found on the ankara dataset, suggest that many predictors yield similar performance.

Generally, the results suggest that our method for choosing $k^\star$ results in penalized trees whose risk remains quite close to that of the unpenalized methods. For example, Table 2 shows that on our ten benchmark regression tasks, penalized CART's estimated risk is always less than 10% higher than CART's. In fact over all $2 \times 4 \times 10 = 80$ possible penalty/criteria/dataset combinations in Tables 2-5, only one has an increase in MSE above 10%. The evidence from classification is similar – in just one case does applying a penalty increase a splitting criterion's holdout misclassification rate by more than 10%. In many cases misclassification rate decreases.

Moreover, the gains in interpretability can be substantial amounting to a "free lunch" of sorts. As an example consider the red wine dataset, where we wish to predict each wine's human-labelled quality score using predictors that measure various aspects of the wine's chemical composition. Figure 10 displays the unpenalized CART tree on the left and the EMA penalized tree on the right. We select $k^\star = .07$ by our usual method with $c = 0.10$. The unpenalized tree uses as many six predictors in a branch, whereas the penalized tree uses only `alcohol` and `sulphates` throughout the entire tree. Moreover, the right tree's fit is easily described as an increasing relationship between `alcohol` and quality for low values of `alcohol` and an increasing relationship between `sulphates` and quality for higher `alcohol` values. The EMA penalty's out-of-bag risk estimate is only 1.8% higher than that of CART's (see Table 2), suggesting we can replace the CART fit with a far more interpretable tree that we can expect to perform essentially just as well on new data.

Figure 10: CART fit to the Red Wine data with and without the EMA Penalty.

(a) Unpenalized
In-sample $R^2 = 0.36$

(b) EMA Penalty ($k^\star = .07$)
In-sample $R^2 = 0.3$

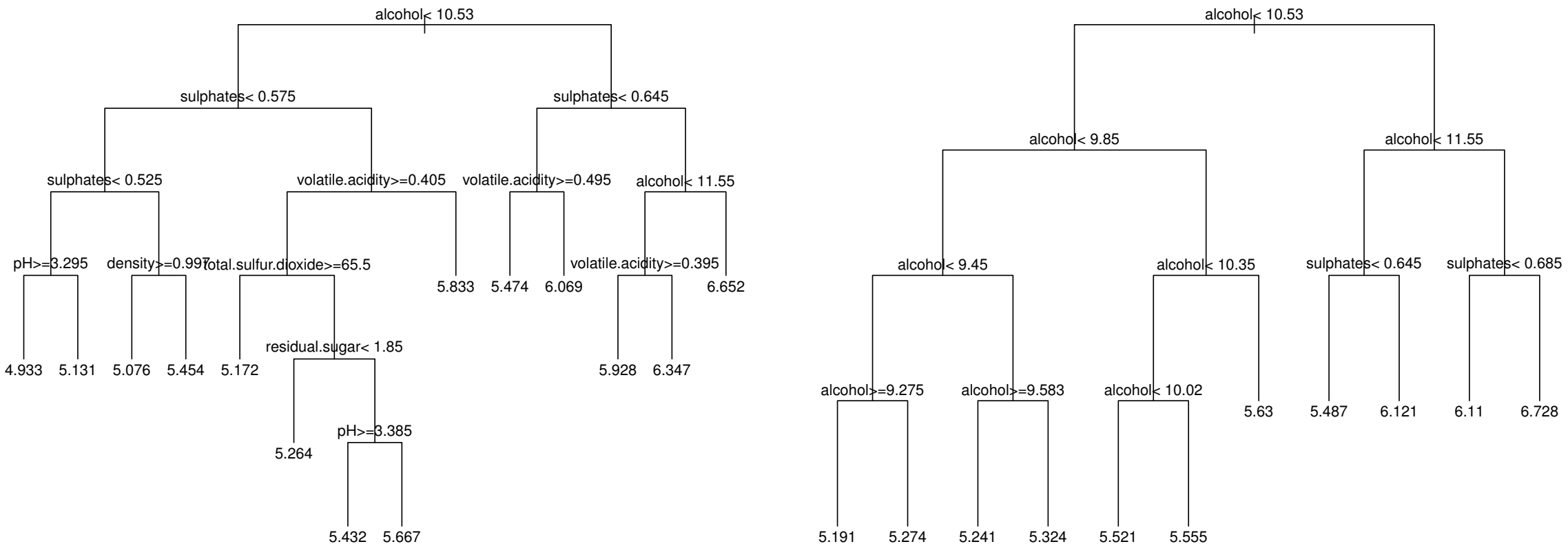

Table 3: OOB performance of penalization methods on **regression** datasets when using the **One-Sided Purity** split criterion. The simulation settings and meaning of the columns follows those in Table 2.

| Dataset | Unpenalized OOB MSE | New Variable Penalty MSE Increase% | New Variable Penalty Average $k^\star$ | EMA Penalty MSE Increase% | EMA Penalty Average $k^\star$ |
|---|---|---|---|---|---|
| boston | 28.68 | 6.8 | 0.12 | 3.5 | 0.24 |
| abalone | 6.07 | 3.3 | 0.47 | 2.4 | 0.71 |
| wine.red | 0.50 | 3.7 | 0.16 | 2.2 | 0.31 |
| wine.white | 0.63 | 5.4 | 0.11 | 4.5 | 0.34 |
| ozone | 25.03 | 2.3 | 0.06 | 1.7 | 0.29 |
| pole | 431.82 | -0.6 | 0.07 | -6.4 | 0.26 |
| triazine | 0.02 | -0.6 | 0.16 | 0.2 | 0.44 |
| ankara | 11.01 | 4.1 | 0.03 | 2.6 | 0.25 |
| baseball | 1009994.68 | -2.3 | 0.10 | -2.2 | 0.28 |
| compactiv | 157.60 | 2.4 | 0.13 | -6.7 | 0.39 |

Table 2: This table displays the OOB performance of the applying the penalization methods to **CART** on 10 **regression** datasets. The two leftmost columns show the average out-of-bag $R^2$ and MSE over 100 bootstrap runs using unpenalized CART. The $R^2$ column is not directly relevant but gives a sense of the difficulty of each problem. The columns titled "MSE Increase%" show the average percentage increase in out-of-bag MSE incurred from applying the penalties with $c = .10$. The "Average $k^\star$" columns show the mean value over the 100 runs of the penalization constant when it is chosen as per (8) with $c = .10$.

| | Unpenalized | | New Variable Penalty | | EMA Penalty | |
|---|---|---|---|---|---|---|
| Dataset | OOB $R^2$ | OOB MSE | MSE Increase% | Average $k^\star$ | MSE Increase% | Average $k^\star$ |
| boston | 0.73 | 22.88 | 1.5 | 0.08 | 0.8 | 0.20 |
| abalone | 0.46 | 5.67 | 5.0 | 0.05 | 4.5 | 0.19 |
| wine.red | 0.29 | 0.47 | 1.8 | 0.04 | 1.8 | 0.12 |
| wine.white | 0.26 | 0.57 | 4.3 | 0.07 | 4.7 | 0.15 |
| ozone | 0.63 | 23.87 | 0.3 | 0.06 | -1.2 | 0.12 |
| pole | 0.78 | 403.37 | 3.0 | 0.08 | 4.1 | 0.18 |
| triazine | 0.06 | 0.02 | -1.7 | 0.05 | -0.9 | 0.14 |
| ankara | 0.96 | 10.16 | 3.3 | 0.11 | 3.4 | 0.30 |
| baseball | 0.57 | 703210.90 | 1.3 | 0.09 | 0.7 | 0.16 |
| compactiv | 0.77 | 78.34 | 4.7 | 0.47 | 5.0 | 0.48 |

Table 4: OOB performance of penalization methods on **regression** datasets when using the **High-Means** split criterion. The simulation settings and meaning of the columns follows those in Table 2.

| | Unpenalized | New Variable Penalty | | EMA Penalty | |
|---|---|---|---|---|---|
| Dataset | OOB MSE | MSE Increase% | Average $k^\star$ | MSE Increase% | Average $k^\star$ |
| boston | 22.19 | 5.9 | 0.03 | 4.5 | 0.11 |
| abalone | 6.10 | 1.1 | 0.99 | 1.1 | 0.99 |
| wine.red | 0.49 | 2.4 | 0.53 | 2.3 | 0.60 |
| wine.white | 0.59 | 6.4 | 0.85 | 6.5 | 0.87 |
| ozone | 24.68 | 2.4 | 0.20 | 3.7 | 0.32 |
| pole | 623.68 | -0.5 | 0.16 | 0.7 | 0.26 |
| triazine | 0.02 | 0.3 | 0.06 | -3.0 | 0.25 |
| ankara | 9.74 | 4.5 | 0.01 | 5.2 | 0.10 |
| baseball | 862508.60 | 6.8 | 0.07 | 4.5 | 0.17 |
| compactiv | 203.87 | 0.7 | 0.04 | -1.4 | 0.10 |

Table 5: OOB performance of penalization methods on **regression** datasets when using the **Low-Means** split criterion. The simulation settings and meaning of the columns follows those in Table 2.

| Dataset | Unpenalized OOB MSE | New Variable Penalty | | EMA Penalty | |
|---|---|---|---|---|---|
| | | MSE Increase% | Average $k^\star$ | MSE Increase% | Average $k^\star$ |
| boston | 23.44 | 11.3 | 0.02 | 8.2 | 0.15 |
| abalone | 6.00 | 4.5 | 0.39 | 3.1 | 0.53 |
| wine.red | 0.48 | 5.9 | 0.02 | 2.8 | 0.13 |
| wine.white | 0.60 | 6.4 | 0.02 | 2.6 | 0.11 |
| ozone | 21.70 | 5.0 | 0.02 | 2.4 | 0.16 |
| pole | 303.06 | 4.6 | 0.05 | 6.1 | 0.14 |
| triazine | 0.02 | 2.7 | 0.08 | -0.0 | 0.25 |
| ankara | 10.36 | 4.5 | 0.01 | 2.7 | 0.08 |
| baseball | 834320.83 | 1.7 | 0.03 | 0.1 | 0.16 |
| compactiv | 81.19 | 5.6 | 0.40 | 3.4 | 0.42 |

Table 6: OOB performance of penalization methods on **classification** datasets when using **CART**. The columns titled "OOB MR" report the average out-of-bag misclassification rate over 100 bootstrap runs. The "Average $k^\star$" columns show the mean value of the penalization constant over the 100 runs when $k^\star$ is chosen as per (8) with $c = .10$.

| Dataset | Unpenalized OOB MR | New Variable Penalty | | EMA Penalty | |
|---|---|---|---|---|---|
| | | OOB MR | Average $k^\star$ | OOB MR | Average $k^\star$ |
| pima | 0.26 | 0.27 | 0.05 | 0.26 | 0.02 |
| breast.cancer | 0.07 | 0.07 | 0.42 | 0.07 | 0.40 |
| bands | 0.33 | 0.33 | 0.98 | 0.33 | 0.98 |
| ionosphere | 0.13 | 0.12 | 0.14 | 0.13 | 0.08 |
| cardio | 0.14 | 0.14 | 0.03 | 0.14 | 0.03 |
| parkinsons | 0.16 | 0.16 | 0.18 | 0.16 | 0.14 |
| glass | 0.36 | 0.37 | 0.01 | 0.37 | 0.01 |
| iris | 0.06 | 0.06 | 0.47 | 0.06 | 0.46 |
| digit.rec | 0.31 | 0.34 | 0.01 | 0.55 | 0.01 |
| waveform1 | 0.28 | 0.29 | 0.03 | 0.29 | 0.01 |

Table 7: OOB performance of penalization methods on **classification** datasets when using **One-Sided Purity**. The simulation settings and meaning of the columns follows those in Table 6.

| Dataset | Unpenalized OOB MR | New Variable Penalty OOB MR | New Variable Penalty Average $k^\star$ | EMA Penalty OOB MR | EMA Penalty Average $k^\star$ |
|---|---|---|---|---|---|
| pima | 0.26 | 0.27 | 0.07 | 0.26 | 0.04 |
| breast.cancer | 0.06 | 0.06 | 0.38 | 0.06 | 0.32 |
| bands | 0.42 | 0.42 | 0.98 | 0.42 | 0.98 |
| ionosphere | 0.18 | 0.18 | 0.13 | 0.18 | 0.08 |
| cardio | 0.14 | 0.14 | 0.03 | 0.14 | 0.03 |
| parkinsons | 0.17 | 0.17 | 0.32 | 0.17 | 0.19 |
| glass | 0.43 | 0.43 | 0.31 | 0.44 | 0.12 |
| iris | 0.10 | 0.11 | 0.48 | 0.11 | 0.47 |
| digit.rec | 0.40 | 0.42 | 0.12 | 0.42 | 0.04 |
| waveform1 | 0.25 | 0.27 | 0.12 | 0.26 | 0.04 |

# 5 Conclusion

This paper describes penalization methods for growing classification and regression trees targeted at settings where interpreting the resultant tree is particularly important. These penalties directly encourage interpretability by controlling the size of the subset of variables used in each branch. By requiring that less interpretable candidate splits decrease the parent node's impurity more than others, penalization allows us to favor interpretability when many splits offer similar improvements. Interestingly, it is the tendency for many splits to offer very similar decreases in impurity – one of CART's perceived disadvantages – that makes this possible.

Using real datasets we show that the penalty functions can indeed result in trees that are substantially easier to explain than their unpenalized counterparts. This observation holds for a variety of splitting criteria and across both classification and regression problems. Further, our study suggests that tuning a penalization parameter to maintain in-sample loss no more than a fraction $c$ of that of the unpenalized procedure's results in generalization error that is almost always within $100c\%$ of the unpenalized method's. That is, in nearly all cases the penalization techniques return a more interpretable fit for very little increase in out-of-sample loss, yielding a "free lunch" of sorts. This raises a number of interesting questions, such as why this is might be the case, what $\mathcal{X}$ designs it is true for, or if further gains can be made by explicitly tuning penalty parameters to minimize holdout loss.

Table 8: Performance of penalization methods on **classification** datasets when using **One-Sided Extremes**. We arbitrarily assign each observed class in the dataset an index $1, .., K$. The columns under "class1" correspond to setting the class of interest to be the first class, and likewise for the second and third. Hence for binary classification problems the third group of columns is blank. When there are more than two classes we report results when the class of interest is set to be the third class in our random ordering. Under each class of interest, the three columns refer to the unpenalized, new variable penalty and EMA procedures, respectively. The first row in a class-method pair reports the mean out-of-bag misclassification rates (100 runs) and the second reports the average $k^\star$ value when $c = 0.10$. For the Pima Indians data, for example, when the class of interest is class1 the average misclassification rate is 0.27 and the average $k^\star$ is 0.66.

| | class1 | | | class2 | | | class3 | | |
|---|---|---|---|---|---|---|---|---|---|
| Dataset | U | NV | EMA | U | NV | EMA | U | NV | EMA |
| pima | 0.27 | 0.27 | 0.27 | 0.28 | 0.28 | 0.28 | | | |
| | | 0.66 | 0.31 | | 0.88 | 0.86 | | | |
| breast.cancer | 0.08 | 0.08 | 0.08 | 0.06 | 0.06 | 0.06 | | | |
| | | 0.40 | 0.29 | | 0.24 | 0.18 | | | |
| bands | 0.32 | 0.32 | 0.32 | 0.44 | 0.44 | 0.44 | | | |
| | | 0.97 | 0.97 | | 0.97 | 0.97 | | | |
| ionosphere | 0.18 | 0.18 | 0.18 | 0.22 | 0.22 | 0.22 | | | |
| | | 0.99 | 0.99 | | 0.59 | 0.50 | | | |
| cardio | 0.16 | 0.17 | 0.17 | 0.17 | 0.17 | 0.18 | 0.17 | 0.17 | 0.17 |
| | | 0.36 | 0.07 | | 0.53 | 0.50 | | 0.87 | 0.86 |
| parkinsons | 0.18 | 0.19 | 0.19 | 0.21 | 0.17 | 0.21 | | | |
| | | 0.71 | 0.60 | | 0.52 | 0.14 | | | |
| glass | 0.51 | 0.49 | 0.50 | 0.53 | 0.53 | 0.54 | 0.61 | 0.53 | 0.54 |
| | | 0.86 | 0.80 | | 0.84 | 0.74 | | 0.87 | 0.86 |
| iris | 0.37 | 0.33 | 0.33 | 0.22 | 0.23 | 0.23 | 0.34 | 0.33 | 0.33 |
| | | 0.94 | 0.92 | | 0.11 | 0.04 | | 0.97 | 0.94 |
| digit.rec | 0.75 | 0.75 | 0.75 | 0.76 | 0.73 | 0.73 | 0.71 | 0.75 | 0.77 |
| | | 0.99 | 0.99 | | 0.99 | 0.99 | | 0.27 | 0.19 |
| waveform1 | 0.40 | 0.43 | 0.45 | 0.40 | 0.42 | 0.43 | 0.39 | 0.42 | 0.42 |
| | | 0.16 | 0.08 | | 0.17 | 0.06 | | 0.28 | 0.13 |

## Acknowledgements

We thank Ed George and Abba Krieger for their insightful comments on this project. We also acknowledge Richard Berk for his thorough reading and substantive edits of an earlier draft.

# A Appendix

## A.1 One-Sided Split Criteria

The one-sided splitting procedures introduced in Buja and Lee (2001) fit into the framework described in Section 2. Whereas all CART techniques use the split criteria defined by Equation 2, all one-sided methods use

$$\theta_{OS}(\mathsf{t}, s) = \phi(\mathsf{t}) - \min\left[\phi(\mathsf{t}_L),\ \phi(\mathsf{t}_R)\right]. \tag{15}$$

Because they ignore the $\phi$ value in one of the child nodes, Buja and Lee (2001) call methods that follow (14) *one-sided*. Buja and Lee (2001) uses two classes of impurity functions $\phi$, resulting in two types of one-sided methods: *one-sided impurity* and *one-sided extremes*. In regression, the former seeks the single child node with lowest MSE whereas the latter seeks the child node with the highest (or lowest) average $y$ value. In classification, one-sided purity uses Gini impurity and one-sided extremes seeks nodes with high sample proportions of a particular class. The impurity functions $\phi$ are defined formally in Table 9.

Table 9: Defintions of One-Sided Impurity Functions

| Procedure | Problem | Impurity Function |
|---|---|---|
| One-sided purity | Regression | $\phi_{osp,R} = \frac{1}{n_\mathrm{t}} \sum_{i \in \mathrm{t}} (y_i - \bar{y}_\mathrm{t})^2$ |
| One-sided purity | Classification | $\phi_{osp,C} = \sum_{k \in K} \hat{p}_{k,\mathrm{t}}(1 - \hat{p}_{k,\mathrm{t}})$ |
| One-sided extremes, high means | Regression | $\phi_{ose,hm} = \bar{y}_\mathrm{t}$ |
| One-sided extremes, low means | Regression | $\phi_{ose,lm} = -\bar{y}_\mathrm{t}$ |
| One-sided extremes | Classification | $\phi_{ose,C} = \hat{p}'_k$ |

Note that the one-sided purity methods using the same impurity function as CART. In regression we use $\phi_R$, the within-node sample variance, and in classification we use $\phi_G$, Gini impurity. In contrast, the one-sided extremes procedures use impurity functions that quantify some aspect of the $y$ values themselves as opposed to their variability. The high (low) means technique finds the single bucket with highest (lowest) sample mean, for example. Note that to use one-sided extremes in a classification setting, the user needs to specify the *class of interest*, denoted $k'$. If we are classifying handwritten digits, for instance, setting $k' =$ "2" means we choose $s^\star$ as to isolate the child node with the highest proportion of observations with $y =$ "2". It is apparent that in general this will be a different split than if we set $k' =$ "3".

## A.2 Gain Function Scaling

As mentioned in section 3, splitting criteria must be adjusted to fit into the penalization framework. Here we give the details of how this is done for each criteria on both classification and regression.

As a motivating example, consider the CART regression tree algorithm. Recall that this algorithm uses the gain function (2) and impurity function (3), resulting in a search for the split yielding the minimal per-observation mean-squared error. If we wish to induce a more interpretable fit by using one of the penalties we must specify the constant $k$. However, the mean-squared error's magnitude varies directly with the level the $y_i$'s, and so penalizing the gain function directly would require us to calibrate $k$ to $y$. We avoid this by scaling $\theta$ by the parent node's impurity, as follows:

$$\theta_{scaled}(\mathfrak{t}, s) = \frac{\theta(\mathfrak{t}, s)}{\phi(\mathfrak{t})} = \frac{\phi(\mathfrak{t}) - \left[\frac{n_{\mathfrak{t}_L}}{n_{\mathfrak{t}}}\phi(\mathfrak{t}_L) + \frac{n_{\mathfrak{t}_R}}{n_{\mathfrak{t}}}\phi(\mathfrak{t}_R)\right]}{\phi(\mathfrak{t})}, \tag{16}$$

where $\phi(\mathfrak{t})$, $\phi(\mathfrak{t}_L)$ and $\phi(\mathfrak{t}_R)$ are the parent node MSE, left daughter MSE and right daugher MSE respectively. Now because the parent node's MSE is constant across all candidate splits we have that

$$\arg\max_{s \in \mathcal{S}} \ \theta_{scaled}(\mathfrak{t}, s) \ = \ \arg\max_{s \in \mathcal{S}} \ \theta(\mathfrak{t}, s),$$

meaning that the optimal split $s^\star$ is invariant to the scaling. Moreover, $\theta_{scaled}$ can be thought of as the fractional improvement in the impurity function, freeing its magnitude from any direct dependence on the level of $y$. Note that if the numerator of (16) is negative it is best not to split, and hence for any feasible $s$ we have $\theta_{scaled}(\mathfrak{t}, s) \in [0, 1]$. Thus we can safely apply a penalty function to $\theta_{scaled}$ using $k \in (0, 1)$. The end result is that splits yielding non-zero penalties (that is, those that are less interpretable) require larger fractional improvements in impurity than those that do not.

There are two features of (16) that make this possible.

- The ordering of $\theta(\mathfrak{t}, s)$ is equivalent to that of $\theta_{scaled}(\mathfrak{t}, s)$ for all $s \in \mathcal{S}$.
- The fact that $0 \leq \theta_{scaled}(\mathfrak{t}, s) \leq 1$ for all feasible $s$.

The first ensures that the scaling does not change the optimal split $s^\star$ and allows us to recover the unpenalized criteria by setting $k = 0$. The second ensures that we can restrict $k \in [0, 1]$. In most situations the unscaled gain function is bounded above by the parent node's impurity function, and so scaling by parent-node impurity suffices. This is the case for CART.

For the one-sided methods such as Buja and Lee (2001), The only case in which this does not work is the high (or low) means one-sided extremes criteria. Table 10 summarizes the specifics.

Table 10: Scaling of Impurity Functions

| Criteria | Impurity Function | Scaling |
|---|---|---|
| CART, (2) | regression | $\phi_R(\mathfrak{t})$ |
| | classification | $\phi_G(\mathfrak{t})$ |
| One-Sided, (14) | purity, regression | $\phi_{osp,R}(\mathfrak{t})$ |
| | purity, classification | $\phi_{osp,C}(\mathfrak{t})$ |
| | extremes, regression; high means | $\max_{i \in \mathfrak{t}}[y_i] - \bar{y}_{\mathfrak{t}}$ |
| | low means | $\bar{y}_{\mathfrak{t}} - \min_{i \in \mathfrak{t}}[y_i]$ |
| | extremes, classification | $\phi_{ose,C}(\mathfrak{t})$ |